\documentstyle[preprint,aps,axodraw]{revtex}
\tightenlines

\def\simgt{\rlap{\lower 3.5 pt\hbox{$\mathchar \sim$}}\raise 1pt \hbox {$>$}}
\def\simlt{\rlap{\lower 3.5 pt\hbox{$\mathchar \sim$}}\raise 1pt \hbox {$<$}}
\newcommand{\nn}{\nonumber}

\newcommand{\ovl}[1]{\overline{#1}}
\newcommand{\wt}[1]{\widetilde{#1}}

\newcommand{\eqn}[1]{(\ref{#1})}

\newcommand{\pslash}{p\kern-1ex /}
\newcommand{\lslash}{l\kern-1ex /}
\newcommand{\Dslash}{{\cal D}\kern-1.5ex /}
\newcommand{\bpsi}{\overline{\psi}}

\newcommand{\msbar}{{\overline {\rm MS}}}
\newcommand{\vev}[1]{\langle #1 \rangle}

\begin{document}


\title{
\vspace{-3.0cm}
\begin{flushright}  
{\normalsize hep-th/9902008}\\
{\normalsize UTHEP-398}\\
\end{flushright}
Perturbative renormalization factors of three- and 
four-quark operators for domain-wall QCD}

\author{$^1$Sinya Aoki, $^1$Taku Izubuchi, $^{2}$Yoshinobu Kuramashi
\thanks{On leave from Institute of Particle and Nuclear Studies,
High Energy Accelerator Research Organization(KEK),
Tsukuba, Ibaraki 305-0801, Japan}
 and $^1$Yusuke Taniguchi}

\address{$^1$Institute of Physics, University of Tsukuba, 
Tsukuba, Ibaraki 305-8571, Japan \\
$^2$Department of Physics, Washington University, 
St. Louis, Missouri 63130, USA \\
}

\date{\today}

\maketitle

\begin{abstract}
Renormalization factors for three- and four-quark operators, which
appear in the low energy effective Lagrangian of 
the proton decay and the weak interactions,  
are perturbatively calculated in domain-wall QCD.
We find that the operators are multiplicatively renormalizable
up to one-loop level 
without mixing with any other operators 
that have different chiral structures.
As an application, we evaluate a renormalization factor
for $B_K$ at the parameters where previous simulations have been performed,
and find one-loop corrections to $B_K$ are 1-5\% in these cases.

\end{abstract}


\narrowtext

\section{Introduction}

Calculation of hadron matrix elements 
of phenomenological interest represents an inevitable
application of lattice QCD.
In the past decade much efforts have been devoted 
for the calculation of three- and four-quark
hadron matrix elements relevant to the proton decay
amplitude and the weak interaction ones 
using the Wilson and the Kogut-Susskind(KS)
quark actions. However, the satisfactorily precise 
measurement of the matrix elements has not been achieved so far
because of the inherent defects in these quark actions:
the explicit chiral symmetry breaking in the Wilson quark action
causes the non-trivial operator mixing between different chiralities 
and for the KS quark it is hard to treat the heavy-light cases
due to the flavor symmetry breaking.

The domain-wall quark formulation in lattice QCD,
which is based on the introduction of many heavy regulator fields, 
was proposed by Shamir\cite{Shamir93,Shamir95} anticipating
superior features over other quark formulations:
no need of the fine tuning to realize the chiral limit
and no restriction for the number of flavors.
Recent simulation results
seem to support the former 
feature non-perturbatively\cite{Blum-Soni,Wingate,Blum}. 
It is also perturbatively shown that 
the massless mode at the tree level still remains
stable against the quantum correction\cite{Aoki-Taniguchi}.
These advantageous features fascinate us to the application of
the domain-wall quark for calculation of 
the three- and four-quark hadron matrix elements.

In order to convert the matrix elements obtained by 
lattice simulations to those defined in some continuum
renormalization scheme(${\it e.g.}, \msbar$),
we must know the renormalization factors 
connecting the lattice operators to the continuum 
counterparts defined in some renormalization scheme.
In this article we make a perturbative calculation of 
the renormalization factors for 
the three- and four-quark operators consisting of physical quark
fields in the domain-wall QCD(DWQCD).
This work is an extension of the previous paper\cite{AIKT98}, in which
we developed a perturbative renormalization procedure for DWQCD 
demonstrating the calculation of 
the renormalization factors for quark wave function, mass 
and bilinear operators.
We focus on whether or not the renormalization of 
the three- and four-quark operators in DWQCD
is free from the notorious operator mixing problem.
In the Wilson case it is well known that 
the mixing problem is not adequately manipulated 
by the perturbation theory,
leading to an ``incorrect'' value for the $B_K$ matrix element.

This paper is organized as follows.
In Sec.~\ref{sec:model} we briefly introduce the DWQCD action and
the Feynman rules relevant for the present calculation to make this
paper self-contained.
In Sec.~\ref{sec:4fermi} our calculational procedure 
of the renormalization factors for the four-quark
operators is described in detail.
We also evaluate the renormalization factors for three-quark operators in
Sec.~\ref{sec:3fermi}. In Secs.~\ref{sec:4fermi} 
and \ref{sec:3fermi} numerical results for one-loop coefficients
of the renormalization factors are given with and
without the mean field improvement.
In Sec.~\ref{sec:bk}, using our results, we analyze a renormalization
factor for $B_K$.
Our conclusions are summarized in Sec.~\ref{sec:concl}.

The physical quantities are expressed in lattice units 
and the lattice spacing $a$ is suppressed unless necessary. 
We take SU($N$) gauge
group with the gauge coupling $g$ and the second Casimir
$C_F = \displaystyle \frac{N^2-1}{2N}$, while
$N=3$ is specified in the numerical calculations.

\section{Action and Feynman rules}
\label{sec:model}

We take the Shamir's domain-wall fermion
action\cite{Shamir93},
\begin{eqnarray}
S_{\rm DW} &=&
\sum_{n} \sum_{s=1}^{N_s} \Biggl[ \frac{1}{2} \sum_\mu
\left( \bpsi(n)_s (-r+\gamma_\mu) U_\mu(n) \psi(n+\mu)_s
+ \bpsi(n)_s (-r-\gamma_\mu) U_\mu^\dagger(n-\mu) \psi(n-\mu)_s \right)
\nn\\&&
+ \frac{1}{2}
\left( \bpsi(n)_s (1+\gamma_5) \psi(n)_{s+1}
+ \bpsi(n)_s (1-\gamma_5) \psi(n)_{s-1} \right)
+ (M-1+4r) \bpsi(n)_s \psi(n)_s \Biggr]
\nn\\&+&
 m \sum_n \left( \bpsi(n)_{N_s} P_{+} \psi(n)_{1}
+ \bpsi(n)_{1} P_{-} \psi(n)_{N_s} \right),
\label{eqn:action}
\end{eqnarray}
where $n$ is a four dimensional space-time coordinate and $s$ is an extra
fifth dimensional or ``flavor'' index,
the Dirac ``mass'' $M$ is a parameter of the theory
which we set $0 < M < 2$ to realize the massless fermion at tree
level, $m$ is a physical quark mass,
and the Wilson parameter is set to $r=-1$.
It is important to notice that we have boundaries for the flavor space;
$1 \le s \le N_s$.
In our one-loop calculation we will take $N_s\to\infty$ limit to avoid
complications arising from the finite $N_s$.
$P_{R/L}$ is a projection matrix $P_{R/L}=(1\pm\gamma_5)/2$.
For the gauge part we employ a standard four dimensional
Wilson plaquette action and assume no gauge interaction along the fifth
dimension.

In the DWQCD the zero mode of domain-wall fermion is
extracted by the ``physical'' quark field defined by the boundary
fermions 
\begin{eqnarray}
q(n) = P_R \psi(n)_1 + P_L \psi(n)_{N_s},
\nn \\
\ovl{q}(n) = \bpsi(n)_{N_s} P_R + \bpsi(n)_1 P_L.
\label{eq:quark}
\end{eqnarray}
We will consider the QCD operators constructed from this quark fields,
since this field has been actually used in the previous simulations.
Moreover our renormalization procedure is based on the Green functions
consisting of only the ``physical'' quark fields, in which
we have found that the renormalization becomes simple\cite{AIKT98}.

Weak coupling perturbation theory is developed by expanding the action
in terms of gauge coupling.
The gluon propagator can be written as
\begin{eqnarray}
G_{\mu\nu}^{AB}(k) =\delta_{\mu\nu}\delta_{AB} \frac{1}{4\sin^2(k/2)+\lambda^2}
\end{eqnarray}
in the Feynman gauge with the infrared cut-off $\lambda^2$,
where $\sin^2(k/2) =\sum_\mu \sin^2(k_\mu/2)$.
Quark-gluon vertices are also identical to those in the $N_s$ flavor
Wilson fermion.
We need only one gluon vertex for our present calculation:
\begin{eqnarray}
V_{1\mu}^A (k,p)_{st}
&=& -i g T^A \{ \gamma_\mu \cos(-k_\mu/2 + p_\mu/2)
  -i r \sin(-k_\mu/2 + p_\mu/2) \} \delta_{st},
\end{eqnarray}
where $k$ and $p$ represent incoming momentum into the vertex
(see Fig.~1 of Ref.\cite{AIKT98}).
$T^A$ $(A=1,\dots,N^2-1)$ is a generator of color 
SU($N$).

The fermion propagator originally takes
$N_s\times N_s$ matrix form in $s$-flavor space.
In the present one-loop calculation, however, we do not
need the whole matrix elements because
we consider Green functions consisting of the
physical quark fields.
The relevant fermion propagators are restricted to
following three types:
\begin{eqnarray}
&&
\vev{q(-p) \ovl{q}(p)} = 
 \frac{-i\gamma_\mu \sin p_\mu + \left(1-W e^{-\alpha}\right) m}
{-\left(1-e^{\alpha}W\right) + m^2 (1-W e^{-\alpha})}
 \equiv S_q(p),
\label{eqn:phys-prop}
\\&&
\vev{q(-p) \bpsi(p,s)} 
=
\frac{1}{F}
\left( i\gamma_\mu \sin p_\mu - m \left(1 -W e^{-\alpha} \right)
\right)
\left( e^{-\alpha (N_s-s)} P_R + e^{-\alpha (s-1)} P_L \right)
\nn\\&&\qquad
+\frac{1}{F} \Bigl[
m \left(i\gamma_\mu \sin p_\mu  -m \left(1-W e^{-\alpha}\right)\right)
- F \Bigr] e^{-\alpha}
\left( e^{-\alpha (s-1)} P_R + e^{-\alpha (N_s-s)} P_L \right),
\\&&
\vev{\psi(-p,s) \ovl{q}(p)} 
=
\frac{1}{F}
\left( e^{-\alpha (N_s-s)} P_L + e^{-\alpha (s-1)} P_R \right)
\left( i\gamma_\mu \sin p_\mu - m \left(1 - W e^{-\alpha} \right)
\right)
\nn\\&&\qquad
+\frac{1}{F}
\left( e^{-\alpha (s-1)} P_L + e^{-\alpha (N_s-s)} P_R \right) e^{-\alpha}
\Bigl[
m \left(i\gamma_\mu \sin p_\mu  -m\left(1- We^{-\alpha}\right) \right)
- F \Bigr]
\end{eqnarray}
with
\begin{eqnarray}
W &=& 1-M -r \sum_\mu (1-\cos p_\mu),
\\
\cosh (\alpha) &=& \frac{1+W^2+\sum_\mu \sin^2 p_\mu}{2|W|},
\label{eq:alpha}
\\
F &=& 1-e^{\alpha} W-m^2 \left(1-W e^{-\alpha}\right),
\label{eq:F}
\end{eqnarray}
where the argument $p$ in the factors $\alpha$ and $W$ is
suppressed. 

In the perturbative calculation of Green functions 
the external quark momenta and masses are assumed to be 
much smaller than the lattice cut-off,
so that we can expand the external quark propagators in terms of them.
We have the following expressions as leading term of the expansion:
\begin{eqnarray}
\langle q\bar q \rangle (p) & = & \frac{1-w_0^2}{i\pslash + (1-w_0^2)m},
\\
\langle q \bar \psi_s \rangle (p) &=&
\langle q\bar q \rangle (p)
\left( w_0^{s-1}P_L + w_0^{N_s-s} P_R\right),
\label{eqn:qpsi}
\\
\langle \psi_s \bar q\rangle (p) &=&
\left( w_0^{s-1}P_R + w_0^{N_s-s} P_L\right) 
\langle q\bar q \rangle (p),
\label{eqn:psiq}
\end{eqnarray}
where $w_0 = 1-M$.

\section{Renormalization factors for four-quark operators}
\label{sec:4fermi}

We consider the following four-quark operators:
\begin{eqnarray}
{\cal O}_\pm & = & \frac{1}{2} \left[
(\bar q_1 \gamma_\mu^L q_2)(\bar q_3 \gamma_\mu^L q_4)
\pm
(\bar q_1 \gamma_\mu^L q_4)(\bar q_3 \gamma_\mu^L q_2) \right], \\
{\cal O}_1 & = & 
-C_F (\bar q_1 \gamma_\mu^L q_2)(\bar q_3 \gamma_\mu^R q_4)
+ (\bar q_1 T^A \gamma_\mu^L q_2)(\bar q_3 T^A \gamma_\mu^R q_4), \\
{\cal O}_2 & = & 
\frac{1}{2N} (\bar q_1 \gamma_\mu^L q_2)(\bar q_3 \gamma_\mu^R q_4)
+ (\bar q_1 T^A \gamma_\mu^L q_2)(\bar q_3 T^A \gamma_\mu^R q_4) ,
\end{eqnarray}
where $\gamma_\mu^{L,R} = \gamma_\mu P_{L,R}$. 
Summation over repeated 
indices such as $\mu$ and $A$ is assumed.
We note that $q_i$ $(i=1,2,3,4)$ are boundary quark fields in DWQCD.
For the convenience of calculation 
we rewrite the above operators as
\begin{eqnarray}
{\cal O}_\pm & = & \frac{1}{2}
\left[1\wt{\otimes}1 \pm 1\wt{\odot}1\right]^{ab;cd} 
\left[(\bar q_1^a \gamma_\mu^L q_2^b)
(\bar q_3^c \gamma_\mu^L q_4^d)\right], \\
{\cal O}_1 & = & \frac{1}{2} 
\left[-N 1\wt{\otimes}1 + 1\wt{\odot}1\right]^{ab;cd}  
\left[(\bar q_1^a \gamma_\mu^L q_2^b)
(\bar q_3^c \gamma_\mu^R q_4^d)\right], \\
{\cal O}_2 & = & \frac{1}{2}
\left[1\wt{\odot}1\right]^{ab;cd}  
\left[(\bar q_1^a \gamma_\mu^L q_2^b)
(\bar q_3^c \gamma_\mu^R q_4^d)\right], 
\end{eqnarray}
where $a,b,c,d$ are color indices, and $\wt{\otimes}$,
$\wt{\odot}$ represent the tensor structures in the color space:
\begin{eqnarray}
\left[1 \wt{\otimes} 1\right]^{ab;cd} & \equiv & \delta_{ab}\delta_{cd}, \\ 
\left[1 \wt{\odot} 1\right]^{ab;cd}  & \equiv & \delta_{ad}\delta_{cb}. 
\end{eqnarray}
To derive these formula,
we have used the Fierz transformation for ${\cal O}_\pm$
and the formula 
\begin{equation}
\sum_A T^A\wt{\otimes} T^A = \frac{1}{2}\left[
-\frac{1}{N} 1\wt{\otimes} 1+1\wt{\odot} 1\right]
\end{equation}
for ${\cal O}_{1,2}$ .

We calculate the following Green function:
\begin{equation}
\langle {\cal O}_\Gamma \rangle_{\alpha\beta;\gamma\delta}^{ij;kl}
\equiv \vev{ {\cal O}_\Gamma (q_1)_{\alpha}^i (\bar q_2)_{\beta}^j
(q_3)_{\gamma}^k (\bar q_4)_{\delta}^l },
\end{equation}
where $\Gamma = \pm, 1,2$. Spinor indices are labeled by
$\alpha,\beta,\gamma,\delta$ and color ones by 
$i,j,k,l$.
Truncating the external quark propagators from
$\langle {\cal O}_\Gamma \rangle$, where we multiply 
$\langle {\cal O}_\Gamma \rangle$ by
${i\pslash_i + (1-w_0^2)m}$, 
we obtain the vertex functions, 
which is written in the following form up to the
one-loop level
\begin{equation}
(1-w_0^2)^4\left(\Lambda_\Gamma\right)_{\alpha\beta;\gamma\delta}^{ij;kl}
=(1-w_0^2)^4\left(\Lambda_\Gamma^{(0)}
+\Lambda_\Gamma^{(1)}\right)_{\alpha\beta;\gamma\delta}^{ij;kl},
\end{equation}
where the superscript $(i)$ refers to the $i$-th loop level and
the trivial factor $(1-w_0^2)^4$ is factored out for the convenience.
We suppress the external momenta $p_i$ 
since the renormalization factor does not depend on them.

The tree level vertex functions $\Lambda_\Gamma^{(0)}$ 
are given by  
\begin{eqnarray}
&\Gamma = \pm, &\qquad
\frac{1}{2}\left[ \gamma_\mu^L 
\otimes \gamma_\mu^L\right]_{\alpha\beta;\gamma\delta}
\left[1\wt{\otimes} 1 \pm 1\wt{\odot} 1\right]^{ij;kl}, 
\label{eq:lambda_pm_0}\\
&\Gamma = 1, &\qquad
\frac{1}{2}\left[ \gamma_\mu^L 
\otimes \gamma_\mu^R \right]_{\alpha\beta;\gamma\delta}
\left[-N 1\wt{\otimes} 1+1\wt{\odot} 1\right]^{ij;kl},
\label{eq:lambda_1_0}\\
&\Gamma = 2, &\qquad
\frac{1}{2}\left[ \gamma_\mu^L 
\otimes \gamma_\mu^R \right]_{\alpha\beta;\gamma\delta}
\left[1\wt{\odot} 1 \right]^{ij;kl}, 
\label{eq:lambda_2_0}
\end{eqnarray}
where $\otimes$ acts on the Dirac spinor space representing 
$[\gamma_X \otimes \gamma_Y ]_{\alpha\beta;\gamma\delta} \equiv
(\gamma_X)_{\alpha\beta}(\gamma_Y)_{\gamma\delta}$. 

The one-loop vertex corrections are 
illustrated by six diagrams in Fig.~1, the sum of which
yields the one-loop level vertex function
\begin{equation} 
\Lambda_\Gamma^{(1)}=
\int_{-\pi}^{\pi}\frac{d^4 k}{(2\pi)^4} 
\left(I_\Gamma^a+,\dots,+I_\Gamma^{c^\prime}\right).
\end{equation}
In order to obtain the expressions for the integrands 
$I_\Gamma^a,\dots,I_\Gamma^{c^\prime}$
we should note that the internal quark propagators 
appearing in the diagrams 
are multiplied by the damping factor which comes from
eqs.\eqn{eqn:qpsi} and \eqn{eqn:psiq}.
The following formula are useful.
\begin{eqnarray}
\langle q \bar \psi_s\rangle \left( w_0^{s-1}P_L + w_0^{N_s-s} P_R\right)
=\left( w_0^{s-1}P_R + w_0^{N_s-s} P_L\right) \langle \psi_s\bar q \rangle 
&=& \frac{i\gamma_\mu \sin p_\mu}{\wt{F}\cdot \wt{F}_0 }\equiv \ovl{G}, 
\label{eq:G_b} \\
\langle q \bar \psi_s\rangle \left( w_0^{s-1}P_R + w_0^{N_s-s} P_L\right)
=\left( w_0^{s-1}P_L + w_0^{N_s-s} P_R\right) \langle \psi_s\bar q \rangle 
&=& -\frac{1}{\wt{F}_0 } \equiv \wt{G}, 
\label{eq:G_t}
\end{eqnarray}
where
$\wt{F}=e^{-\alpha}-W$ and $\wt{F}_0 = e^\alpha-w_0$.
Here we set $m=p_i=0$ for the internal propagator.

The contribution from Fig.~1a takes the form 
\begin{eqnarray}
I_\Gamma^a &=& \frac{1}{2} J_a^{AB}
\left\{\ovl{V}_\mu(k) \ovl{G}(k) + \wt{V}_\mu (k)\wt{G}(k)
\right\}
\Gamma_X
\left\{\ovl{G}(k) \ovl{V}_\nu(k) + \wt{G}(k) \wt{V}_\nu(k)\right\}
\otimes \Gamma_Y
\, G_{\mu\nu}^{AB}(k),
\end{eqnarray}
where $\Gamma_X = \gamma_\mu^L$, 
$\Gamma_Y = \gamma_\mu^L$ or $\gamma_\mu^R$, and
the interaction vertices are
\begin{equation}
\ovl{V}_\mu = -i g \gamma_\mu \cos(k_\mu/2),\quad
\wt{V}_\mu = -rg \sin(k_\mu/2).
\label{eq:V_bt}
\end{equation}
The color factors are represented by $J_a^{AB}$, which are
listed in Table~\ref{tab:color}.
In a similar way the contributions from Fig.~1b and Fig.~1c are
given by
\begin{eqnarray}
I_\Gamma^b &=& \frac{1}{2} J_b^{AB}
\left\{\ovl{V}_\mu(k) \ovl{G}(k) + \wt{V}_\mu(k) \wt{G}(k)\right\}
\Gamma_X\otimes
\left\{\ovl{V}_\nu (-k) \ovl{G}(-k) + \wt{V}_\nu (-k)\wt{G}(-k)\right\}
\Gamma_Y
\, G_{\mu\nu}^{AB}(k), \\
I_\Gamma^c &=& \frac{1}{2} J_c^{AB}
\left\{\ovl{V}_\mu(k) \ovl{G}(k) + \wt{V}_\mu (k) \wt{G}(k)\right\}
\Gamma_X\otimes
\Gamma_Y
\left\{\ovl{G}(k)\ovl{V}_\nu(k)  + \wt{G}(k) \wt{V}_\nu(k) \right\}
\, G_{\mu\nu}^{AB}(k).
\end{eqnarray}
After a little algebra the expressions of $I_\Gamma^{a,b,c}$ are reduced to 
\begin{eqnarray}
I_\Gamma^a &=&  \frac{1}{2} g^2 J_a^{AA} 
K \left[ T + A_{VA}\right]\left[\Gamma_X\otimes\Gamma_Y\right], 
\label{eq:vtx_a}\\
I_\Gamma^b &=& - \frac{1}{2} g^2 J_b^{AA} 
K \left[T \Gamma_X\otimes \Gamma_Y +
\cos^2(k_\mu/2) \sin^2 k_\alpha (\gamma_\mu\gamma_\alpha\Gamma_X)
\otimes (\gamma_\mu\gamma_\alpha\Gamma_Y) \right], 
\label{eq:vtx_b}\\
I_\Gamma^c &=&  \frac{1}{2} g^2 J_c^{AA} 
K \left[T \Gamma_X \otimes \Gamma_Y +
\cos^2(k_\mu/2) \sin^2 k_\alpha (\gamma_\mu\gamma_\alpha\Gamma_X)
\otimes (\Gamma_Y\gamma_\alpha\gamma_\mu) \right],
\label{eq:vtx_c}
\end{eqnarray}
where
\begin{eqnarray}
&&
K =\displaystyle \frac{1}{\wt{F}^2\wt{F}_0^2 (4\sin^2(k/2)+\lambda^2)},
\label{eq:K}\\&&
T=r^2 \sin^2(k/2) \wt{F}^2+r \sin^2 k \wt{F},
\label{eq:T}\\&&
A_{VA}= \sum_\mu \cos^2(k_\mu/2) \sin^2 k_\mu.
\label{eq:A_VA}
\end{eqnarray}

In order to rewrite the second term of $I_\Gamma^{b,c}$ 
we apply the Fierz transformation:
\begin{eqnarray}
(\gamma_\mu\gamma_\alpha\gamma_\nu^L)\otimes 
(\gamma_\mu\gamma_\alpha\gamma_\nu^L) &=&
-\gamma_\nu^L\odot\gamma_\nu^L
 = \gamma_\nu^L\otimes\gamma_\nu^L,
\\
(\gamma_\mu\gamma_\alpha\gamma_\nu^L)\otimes 
(\gamma_\nu^L\gamma_\alpha\gamma_\mu) &=&
-\gamma_\nu^L\odot\gamma_\nu^L (1-2\delta_{\alpha\nu})(1-2\delta_{\mu\nu}) 
\\&=&
-\gamma_\nu^L\otimes\gamma_\nu^L (1-2\delta_{\alpha\nu})
(1-2\delta_{\mu\nu})
+2\gamma_\nu^L\otimes\gamma_\nu^L \delta_{\alpha\mu},
\\
(\gamma_\mu\gamma_\alpha\gamma_\nu^L)\otimes 
(\gamma_\mu\gamma_\alpha\gamma_\nu^R) &=&
2P_R\odot P_L \delta_{\mu\alpha}
 = \gamma_\nu^L \otimes \gamma_\nu^R \delta_{\mu\alpha},
\label{eq:lr1_fierz}\\
(\gamma_\mu\gamma_\alpha\gamma_\nu^L)\otimes 
(\gamma_\nu^R\gamma_\alpha\gamma_\mu) &=&
2P_R\odot P_L
 = \gamma_\nu^L \otimes \gamma_\nu^R .
\label{eq:lr2_fierz}
\end{eqnarray}
where $[\gamma_X \odot \gamma_Y ]_{\alpha\beta;\gamma\delta} \equiv
(\gamma_X)_{\alpha\delta}(\gamma_Y)_{\gamma\beta}$,
and summation over $\nu$ is taken.
The Fierz transformation is again used for the second equality.
We omit the tensor term in eq.(\ref{eq:lr1_fierz}) since it vanishes 
in the integral.

Choosing $\Gamma_{X,Y} = \gamma_\mu^L$
in eqs.(\ref{eq:vtx_a}), (\ref{eq:vtx_b}) and (\ref{eq:vtx_c})
we first consider the case of ${\cal O}_{\pm}$. 
After simplifying the expressions of the color factors
we obtain
\begin{eqnarray}
I_\pm^a &=&  \frac{1}{2} g^2 
        K (T+A_{VA})\left[\gamma_\nu^L\otimes\gamma_\nu^L\right]
        \left[ (C_F\pm \frac{1}{2}) 1\wt{\otimes}1
               \mp \frac{1}{2N} 1\wt{\odot}1 \right], \\
I_\pm^b &=& -  \frac{1}{2} g^2 
        K (T+A_{SP})\left[\gamma_\nu^L\otimes\gamma_\nu^L\right]
        \left[ (-\frac{1}{2N}\pm \frac{1}{2}) 1\wt{\otimes}1
               +(\frac{1}{2}\mp \frac{1}{2N}) 1\wt{\odot}1 \right], \\
I_\pm^c &=&  \frac{1}{2} g^2 
        K (T+A_{VA})\left[\gamma_\nu^L\otimes\gamma_\nu^L\right]
        \left[ -\frac{1}{2N} 1\wt{\otimes}1
               +(\frac{1}{2}\pm C_F) 1\wt{\odot}1 \right]. 
\end{eqnarray}
The total contribution becomes
\begin{eqnarray}
I_+^a+I_+^b+I_+^c &=&  \frac{1}{2} g^2 K 
\left[\gamma_\nu^L\otimes\gamma_\nu^L\right]_{\alpha\beta;\gamma\delta} 
\left[1\wt{\otimes}1+1\wt{\odot}1\right]^{ij;kl} \nonumber \\
&\times &
\{ T C_F + A_{VA}(C_F +\frac{1}{2} - \frac{1}{2N})
-A_{SP}(\frac{1}{2} - \frac{1}{2N}) \},
\label{eq:sumP}
\end{eqnarray}
and
\begin{eqnarray}
I_-^a+I_-^b+I_-^c &=&  \frac{1}{2} g^2 K 
\left[\gamma_\nu^L\otimes\gamma_\nu^L\right]_{\alpha\beta;\gamma\delta} 
\left[1\wt{\otimes}1-1\wt{\odot}1\right]^{ij;kl} \nonumber \\
&\times &
\{ T C_F + A_{VA}(C_F -\frac{1}{2} - \frac{1}{2N})
+A_{SP}(\frac{1}{2} + \frac{1}{2N}) \},
\label{eq:sumM}
\end{eqnarray}
where 
\begin{equation}
A_{SP}=\cos^2(k/2)\cdot \sin^2 k.
\label{eq:A_SP}
\end{equation}
We should note that the other three contributions 
$I_\pm^{a^\prime,b^\prime,c^\prime}$  from Fig.~1$a^\prime$, 
1$b^\prime$ and 1$c^\prime$ are
equal to $I_\Gamma^{a,b,c}$ respectively,
therefore
the factors $1/2$ in eqs.~(\ref{eq:sumP}) and (\ref{eq:sumM})
disappear in the total contributions of all.
 
Comparing the one-loop results to the tree level ones 
we obtain
\begin{eqnarray}
\Lambda_+ &=&
\left[ 1 + g^2 \frac{N-1}{N}\left\{ 
<T> (N+1)+<A_{VA}>(N+2)-<A_{SP}> \right\}\right]
\Lambda_+^{(0)}, 
\label{eq:1loop_+}\\
\Lambda_- &=&
\left[ 1 + g^2 \frac{N+1}{N}\left\{ 
<T> (N-1)+<A_{VA}>(N-2)+<A_{SP}>\right\}\right]
\Lambda_-^{(0)}
\label{eq:1loop_-} 
\end{eqnarray}
with
\begin{equation}
< X > = \int_{-\pi}^{\pi}\frac{d^4 k}{(2\pi)^4} K(k) X(k)
\label{eq:integ_X}
\end{equation}
for $X = T, A_{VA}, A_{SP}$.
We remark that $C_F< T+A_{VA} >$ and $C_F< T+A_{SP} >$ correspond
to the one-loop vertex corrections to the (axial) vector current 
and the (pseudo) scalar density which are expressed as
$(T_{VA}-1)$ and $(T_{SP}-1)$ in Ref.~\cite{AIKT98}.
The expressions of eqs.(\ref{eq:1loop_+}) and (\ref{eq:1loop_-}) 
show an important
property of ${\cal O}_{\pm}$ in the DWQCD formalism: 
the one-loop vertex corrections are multiplicative.
This is contrary to the Wilson case, in which the mixing operators
with different chiralities appears at the one-loop level\cite{pt_w4}. 

We next turn to the case of ${\cal O}_{1,2}$.
For ${\cal O}_1$ the vertex corrections 
of eqs.(\ref{eq:vtx_a}), (\ref{eq:vtx_b}) and (\ref{eq:vtx_c})
with $\Gamma_X = \gamma_\mu^L$ and $\Gamma_Y = \gamma_\mu^R$
are written as 
\begin{eqnarray}
I_1^a &=&  \frac{1}{2} g^2 K (T+A_{VA})
          \left[\gamma_\nu^L\otimes\gamma_\nu^R\right] 
          \left[ (-N C_F+ \frac{1}{2}) 1\wt{\otimes}1
                 - \frac{1}{2N} 1\wt{\odot}1 \right], \\
I_1^b &=& - \frac{1}{2} g^2 K (T+A_{VA})
          \left[\gamma_\nu^L\otimes\gamma_\nu^R\right] 
          \left[ (\frac{1}{2}+ \frac{1}{2}) 1\wt{\otimes}1
                 +(-\frac{N}{2}- \frac{1}{2N}) 1\wt{\odot}1 \right], \\
I_1^c &=&  \frac{1}{2} g^2 K (T+A_{SP})
          \left[\gamma_\nu^L\otimes\gamma_\nu^R\right]
          \left[ \frac{1}{2} 1\wt{\otimes}1
                 +(-\frac{N}{2}+ C_F) 1\wt{\odot}1 \right]. 
\end{eqnarray}
The total contribution including those from Fig.~$1a^\prime$, $1b^\prime$ 
and $1c^\prime$ is given by
\begin{eqnarray}
2(I_1^a+I_1^b+I_1^c) &=& 2\frac{1}{2} g^2 
K \left[\gamma_\nu^L\otimes\gamma_\nu^R\right]_{\alpha\beta;\gamma\delta} 
\left[- N 1\wt{\otimes} 1 +1\wt{\odot} 1\right]^{ij;kl} \nonumber
\\&\times &
\left[ T C_F + A_{VA}\frac{N}{2}-A_{SP}\frac{1}{2N} \right].
\end{eqnarray}
Using the tree level result in eq.(\ref{eq:lambda_1_0})
the vertex function up to the one-loop level is expressed as 
\begin{eqnarray}
\Lambda_1 &=&
\left[ 1 + g^2 \frac{1}{N}\left\{ 
<T> (N^2-1)+<A_{VA}>N^2-<A_{SP}>\right\}\right]
\Lambda_1^{(0)}.
\end{eqnarray}
This result shows that the operator ${\cal O}_1$ is 
multiplicatively renormalizable in DWQCD, which 
is in contrast with the Wilson case\cite{pt_w4}. 

In a similar way we write the vertex corrections
for ${\cal O}_2$.
\begin{eqnarray}
I_2^a &=&  \frac{1}{2} g^2 K (T+A_{VA})
          \left[\gamma_\nu^L\otimes\gamma_\nu^R\right] 
          \left[ \frac{1}{2} 1\wt{\otimes}1
                 - \frac{1}{2N} 1\wt{\odot}1 \right], \\
I_2^b &=& - \frac{1}{2} g^2 K (T+A_{VA})
          \left[\gamma_\nu^L\otimes\gamma_\nu^R\right] 
          \left[ \frac{1}{2} 1\wt{\otimes}1
                 - \frac{1}{2N} 1\wt{\odot}1 \right], \\
I_2^c &=& \frac{1}{2} g^2 K (T+A_{SP})
          \left[\gamma_\nu^L\otimes\gamma_\nu^R\right] 
          \left[ C_F 1\wt{\odot}1 \right]. 
\end{eqnarray}
The total contribution including those from 
Fig.~$1a^\prime$, $1b^\prime$ 
and $1c^\prime$ becomes
\begin{eqnarray}
2(I_2^a+I_2^b+I_2^c) &=& 2\frac{1}{2}g^2 
K \left[\gamma_\nu^L\otimes\gamma_\nu^R\right]_{\alpha\beta;\gamma\delta} 
\left[1\wt{\odot} 1 \right]^{ij;kl} \nonumber \\
&\times &
C_F\left[ T + A_{SP}\right],
\end{eqnarray}
which leads to
\begin{eqnarray}
\Lambda_2 &=&
\left[ 1 + g^2 \frac{N^2-1}{N}\left\{ <T> +<A_{SP}>\right\}\right]
\Lambda_2^{(0)}.
\end{eqnarray}
We again find that the vertex correction is multiplicative 
up to the one-loop level as opposed to the Wilson case\cite{pt_w4}.

The contribution from the fermion self-energy has already been evaluated
\cite{Aoki-Taniguchi,AIKT98} and the total lattice renormalization
factor is now obtained: 
\begin{equation}
Z_\Gamma^{lat} = (1-w_0^2)^2 Z_w^2 Z_2^2 V_\Gamma,
\end{equation}
where
\begin{eqnarray}
Z_2 &=& 1 + \frac{g^2}{16\pi^2} C_F \left[ \log (\lambda a)^2
 + \Sigma_1\right],
\\
V_\Gamma &=& 1 + \frac{g^2}{16\pi^2} \left[ -\delta_\Gamma \log (\lambda a)^2
+ v_\Gamma \right],
\end{eqnarray}
\begin{eqnarray}
v_+ &=&\frac{16\pi^2(N-1)}{N}\left[ <T> (N+1)+<<A_{VA}>>(N+2)-<<A_{SP}>>
\right] +\delta_+ \log \pi^2,
 \\
v_- &=&\frac{16\pi^2(N+1)}{N}\left[ <T> (N-1)+<<A_{VA}>>(N-2)+<<A_{SP}>>
\right] +\delta_- \log \pi^2,
 \\
v_1 &=&\frac{16\pi^2}{N}\left[ <T> (N^2-1)+<<A_{VA}>>N^2-<<A_{SP}>>
\right] +\delta_1 \log \pi^2,
 \\
v_2 &=&\frac{16\pi^2(N^2-1)}{N}\left[ <T> + <<A_{SP}>>
\right] +\delta_2 \log \pi^2 
\end{eqnarray}
with
\[
\delta_\Gamma =
\left\{
\begin{array}{ll}
\displaystyle \frac{(N-1)(N-2)}{N} & \qquad \Gamma = + \\
& \\
\displaystyle \frac{(N+1)(N+2)}{N} & \qquad \Gamma = - \\
& \\
\displaystyle \frac{(N+2)(N-2)}{N} & \qquad \Gamma = 1 \\
& \\
\displaystyle \frac{4(N+1)(N-1)}{N} & \qquad \Gamma = 2 
\end{array}
\right.
\]
The infrared singularity of $< A_X>$ is subtracted as
\[
<< A_X >> = \int_{-\pi}^{\pi}\frac{d^4k}{(2\pi)^4} 
\left[K(k) A_X(k) - c_X \frac{1}{(k^2)^2}\theta(\pi^2-k^2)\right]
\]
with $c_{SP}=4$ and $c_{VA}=1$.

Numerical values of $v_\Gamma$ are evaluated by two independent methods.
In one method the momentum integration is performed by a mode sum for a
periodic box of a size $L^4$ after transforming the momentum variable through
$k_\mu = q_\mu -\sin q_\mu$. We employ the size $L=64$ for integrals.
In the other method the momentum integration is carried out by
the the Monte Carlo integration routine VEGAS,
using 20 samples of 1000000 points each. We find that both results agree
very well. Numerical values of $v_\Gamma$ are presented 
in Table~\ref{tab:4fermi} as a function of $M$.

We have to also calculate the corresponding continuum wave-function 
renormalization factor and vertex corrections in the
$\msbar$ scheme employing the same gauge and the same
infrared regulator as the lattice case.
For the present calculation 
it seems preferable to choose 
Dimensional Reduction(DRED) as the
ultraviolet regularization, in which the loop momenta
of the Feynman integrals are defined in $D<4$ dimensions
while keeping the Dirac matrices in four dimensions.  
In the DRED scheme
we can use the same calculational techniques
for the vertex corrections as the lattice case 
thanks to applicability of the Fierz transformation for the  
Dirac matrices.
For the wave-function renormalization factor
a simple calculation gives
\begin{equation}
Z_2^{\msbar} = 1+ \frac{g^2}{16\pi^2}C_F \left[
\log (\lambda/\mu)^2 -1/2\right],
\end{equation}
where $\mu$ is a renormalization scale.
This result leads to $\Sigma_1^{\msbar}=-1/2$.
For the vertex corrections we obtain
\begin{equation}
V_\Gamma^{\msbar} = 1+\frac{g^2}{16\pi^2} \delta_\Gamma \left[
-\log (\lambda/\mu)^2 +1\right],
\end{equation}
giving $v_\Gamma^{\msbar}=\delta_\Gamma$.
Here we should remark that the one-loop vertex corrections yield 
the evanescent operators which vanish in $D=4$ 
for the DRED scheme\cite{BW}. It is meaningless to
give results without mentioning the definition of evanescent
operators, because the constant terms at the one-loop level
depend on the definition of the evanescent operators.
Our choice is as follows:
\begin{eqnarray}
E^{\rm DRED}_{\pm}=&
{\bar \delta}_{\mu\nu}\gamma_\mu(1-\gamma_5)
\otimes\gamma_\nu(1-\gamma_5)
-\frac{D}{4}\gamma_\mu(1-\gamma_5)\otimes\gamma_\mu(1-\gamma_5), \\
E^{\rm DRED}_{1,2}=&
{\bar \delta}_{\mu\nu}\gamma_\mu(1-\gamma_5)
\otimes\gamma_\nu(1+\gamma_5)
-\frac{D}{4}\gamma_\mu(1-\gamma_5)\otimes\gamma_\mu(1+\gamma_5),
\end{eqnarray}  
where ${\bar \delta}_{\mu\nu}$ is the $D$-dimensional
metric tensor which emerges inevitably in the evaluation
of the Feynman integrals.
 
Combining these results with the previous lattice ones we obtain
\begin{eqnarray}
{\cal O}_\Gamma^{\msbar}(\mu) & = &\frac{1}{(1-w_0^2)^2 Z_w^2}
Z_\Gamma (\mu a ) {\cal O}_\Gamma^{lat} (1/a),
\end{eqnarray}
where
\begin{eqnarray}
Z_\Gamma (\mu a) &=& \frac{ (Z_2^{\msbar})^2 V_\Gamma^{\msbar}}
{(Z_2)^2 V_\Gamma} \nonumber \\
&=& 1 + \frac{g^2}{16\pi^2}\left[
(\delta_\Gamma - 2 C_F)\log (\mu a)^2 + z_\Gamma \right],
 \label{eq:4fermi}
\\
z_\Gamma &=& v_\Gamma^{\msbar} - v_\Gamma + 2 C_F\{\Sigma_1^{\msbar}
-\Sigma_1\} .
\end{eqnarray}
Numerical values of $z_\Gamma$ are given in 
Table~\ref{tab:total} and
the results for the mean-field improved one, $z_\Gamma^{MF}$ ,
are also given in Table~\ref{tab:totalMF}.

Although the results for the DRED scheme are presented here,
it is an easy task to obtain those for the Naive Dimensional
Regularization(NDR) scheme. 
In Appendix B we summarize the finite parts of the wave-function 
renormalization factor and vertex corrections in the NDR scheme.

\section{Renormalization factors for three-quark operators}
\label{sec:3fermi}

The three-quark operators relevant to the proton decay
amplitude are given by
\begin{equation}
\left({\cal O}_{PD}\right)_\delta =\varepsilon^{abc} 
\left((\bar q^C_1)^a \Gamma_X (q_2)^b\right) (\Gamma_Y (q_3)^c)_\delta,
\label{eq:O_PD}
\end{equation}
where $\bar q^C = -q^T C^{-1}$ with $C=\gamma_0\gamma_2$ is a
charge conjugated field of $q$ and 
$\Gamma_X\otimes\Gamma_Y = P_R\otimes P_R, P_R\otimes P_L, 
P_L\otimes P_R, P_L\otimes P_L$. 
The summation over repeated color indices $a,b,c$ is assumed.
We should note that the domain-wall fermion action \eqn{eqn:action}
is transformed identically into that with
the conjugated field by using transpose and matrix $C$.
The resultant action and Feynman rules for the conjugated field is
obtained by the replacement that
\begin{eqnarray}
ig T^A &\rightarrow & -ig (T^A)^T ,
\end{eqnarray}
where the superscript $T$ means the transposed matrix.

In order to evaluate the vertex corrections
we consider the following Green function:
\begin{equation}
\langle \left({\cal O}_{PD}\right)_\delta 
\rangle_{\alpha\beta\gamma}^{ijk}
\equiv \vev{ \left({\cal O}_{PD}\right)_\delta 
(q^C_1)_\alpha^i(\bar q_2)_\beta^j
(\bar q_3)_\gamma^k },
\end{equation}
where $\alpha,\beta,\gamma$ and $i,j,k$ are spinor and color indices
respectively.
Truncating the external quark propagators of
$\langle {\cal O}_{PD}\rangle$
we obtain the vertex function
\begin{equation}
(1-w_0^2)^3\left(\Lambda_{PD}\right)_{\alpha\beta;\delta\gamma}^{ijk}
=(1-w_0^2)^3\left(\Lambda_\Gamma^{(0)}
+\Lambda_\Gamma^{(1)}\right)_{\alpha\beta;\delta\gamma}^{ijk},
\end{equation}
where the trivial factor $(1-w_0^2)^3$ 
is factored out for the convenience.
We suppress the external momenta $p_i$ 
since the renormalization factor does not depend on them.

At the tree level the vertex function takes the form
\begin{equation}
\Lambda_\Gamma^{(0)}=\varepsilon^{ijk} \left[ \Gamma_X \otimes \Gamma_Y 
\right]_{\alpha\beta;\delta\gamma},
\label{eq:PD_tree}
\end{equation}
where $[\Gamma_X \otimes \Gamma_Y]_{\alpha\beta;\delta\gamma} \equiv
(\Gamma_X)_{\alpha\beta}(\Gamma_Y)_{\delta\gamma}$.

The one-loop vertex corrections are shown in Figs.~2a, 2b and 2c,
the sum of which gives
the one-loop level vertex function
\begin{equation} 
\Lambda_{PD}^{(1)}=
\int_{-\pi}^{\pi}\frac{d^4 k}{(2\pi)^4} 
\left(I_{PD}^a+I_{PD}^b+I_{PD}^{c}\right).
\end{equation}
Using the notations in eqs.(\ref{eq:G_b}), (\ref{eq:G_t}) 
and (\ref{eq:V_bt}) the integrands 
$I_{PD}^{a,b,c}$ are written as follows:
\begin{eqnarray}
I_{PD}^a &=& \varepsilon^{abk}(-T^A)^T_{ia}T^B_{bj} \nn \\
&&\times \left\{\ovl{V}_\mu(k) \ovl{G}(k) 
+ \wt{V}_\mu (k)\wt{G}(k)\right\}
\Gamma_X
\left\{\ovl{G}(k) \ovl{V}_\nu(k) 
+ \wt{G}(k) \wt{V}_\nu(k)\right\}
\otimes \Gamma_Y
G_{\mu\nu}^{AB}(k), \\
I_{PD}^b &=& \varepsilon^{ibc}T^A_{bj}T^B_{ck} \nn \\
&&\times \Gamma_X\left\{\ovl{G}(k) \ovl{V}_\nu(k) 
+ \wt{G}(k) \wt{V}_\nu(k)\right\}
\otimes\Gamma_Y\left\{\ovl{G}(-k) \ovl{V}_\nu(-k) 
+ \wt{G}(-k) \wt{V}_\nu(-k)\right\}
G_{\mu\nu}^{AB}(k), \\
I_{PD}^c &=& \varepsilon^{ajc}(-T^A)^T_{ia}T^B_{ck} \nn \\
&&\times \left\{\ovl{V}_\mu(k) \ovl{G}(k) + \wt{V}_\mu (k)
\wt{G}(k)\right\}
\Gamma_X\otimes
\Gamma_Y
\left\{\ovl{G}(k)\ovl{V}_\nu(k)  
+ \wt{G}(k)\wt{V}_\nu(k) \right\}
G_{\mu\nu}^{AB}(k).
\end{eqnarray}
A little algebra yields
\begin{eqnarray}
I_{PD}^a &=& g^2\frac{N+1}{2N}\varepsilon^{ijk} 
K \left[ T + A_{SP}\right]\left[\Gamma_X\otimes\Gamma_Y\right] , \\
I_{PD}^b &=& g^2\frac{N+1}{2N}\varepsilon^{ijk}  
K \left[T (\Gamma_X\otimes\Gamma_Y) +
\cos^2 (k_\mu/2) \sin^2 k_\alpha  (\Gamma_X\gamma_\alpha\gamma_\mu)
\otimes (\Gamma_Y\gamma_\alpha\gamma_\mu) \right], \\
I_{PD}^c &=& g^2\frac{N+1}{2N}\varepsilon^{ijk} 
K \left[T (\Gamma_X \otimes\Gamma_Y) +
\cos^2 (k_\mu/2) \sin^2 k_\alpha (\gamma_\mu\gamma_\alpha\Gamma_X)
\otimes (\Gamma_Y\gamma_\alpha\gamma_\mu) \right],
\end{eqnarray}
where $K$, $T$ and $A_{SP}$ are given 
in eqs.(\ref{eq:K}), (\ref{eq:T}) and (\ref{eq:A_SP}). 
It is noted that a sum of $I_{PD}^b$ and $I_{PD}^c$ becomes
\begin{eqnarray}
& &g^2\frac{N+1}{2N}\varepsilon^{ijk} 
K \left[2T (\Gamma_X \otimes\Gamma_Y) +
\cos^2 (k_\mu/2) \sin^2 k_\alpha (\gamma_\mu\gamma_\alpha\Gamma_X+
\Gamma_X\gamma_\alpha\gamma_\mu)
\otimes (\Gamma_Y\gamma_\alpha\gamma_\mu) \right] \nonumber \\
& = &
g^2\frac{N+1}{2N}\varepsilon^{ijk} 
K \left[2T (\Gamma_X \otimes\Gamma_Y) +
\cos^2 (k_\mu/2) \sin^2 k_\alpha (\{\gamma_\mu\gamma_\alpha
+\gamma_\alpha\gamma_\mu \}\Gamma_X)
\otimes (\Gamma_Y\gamma_\alpha\gamma_\mu) \right ]
\end{eqnarray}
for $\Gamma_X = P_R$ or $P_L$, therefore
no Fierz transformation is necessary to simplify the spinor structure
of the total contribution.
Finally
we obtain 
\begin{eqnarray}
I_{PD}^a+I_{PD}^b+I_{PD}^c &=& g^2\frac{N+1}{2N}\varepsilon^{ijk}
K \left[\Gamma_X\otimes\Gamma_Y\right] 
\left[ 3T + A_{SP} + 2A_{VA} \right].
\end{eqnarray}
Compared with the tree level result of eq.(\ref{eq:PD_tree}) we find
that the vertex correction is multiplicative 
up to the one-loop level:
\begin{equation}
\Lambda_{PD} =
\left[ 1 + g^2 \frac{N+1}{2N}\left\{ 
3<T> + <A_{SP}> + 2<A_{VA}> \right\}\right]
\Lambda_{PD}^{(0)},
\end{equation}
where $<X>$ $(X = T, A_{VA}, A_{SP})$ are defined 
in eq.(\ref{eq:integ_X}).
We remark that in the Wilson case ${\cal O}_{PD}$ mixes
with other operators which have different chiral structures
under renormalization\cite{pt_w3}.

Taking account of the contribution of the wave function 
the lattice renormalization factor for ${\cal O}_{PD}$ 
is expressed as
\begin{equation}
Z_{PD}^{lat} = (1-w_0^2)^{3/2}Z_w^{3/2} Z_2^{3/2}V_{PD},
\end{equation}
where
\begin{eqnarray}
V_{PD} &=& 1+ \frac{g^2}{16\pi^2}\left[-\delta_{PD}\log (\lambda a)^2 +
v_{PD}\right],
\end{eqnarray}
\begin{eqnarray}
v_{PD} &=& \frac{16\pi^2 (N+1)}{2N}\left[ 3 <T> + <<A_{SP}>> +
2 <<A_{VA}>> \right]+\delta_{PD}\log\pi^2
\end{eqnarray}
with
$\delta_{PD}=\displaystyle\frac{6(N+1)}{2N}$.
Numerical values for $v_{PD}$, evaluated as before, 
are given in Table~\ref{tab:4fermi} as a function of $M$.

The corresponding continuum renormalization factors
in the $\msbar$ scheme are calculated 
employing the DRED scheme as the regularization 
in the Feynman gauge with the fictitious gluon mass $\lambda$.  
The vertex correction for ${\cal O}_{PD}$ is
\begin{eqnarray}
V_{PD}^{\msbar} &=& 1+ \frac{g^2}{16\pi^2}\delta_{PD}
\left[-\log (\lambda/\mu)^2 +1 \right],
\end{eqnarray}
giving $v_{PD}^{\msbar}=\delta_{PD}$.
We remark that in this case the evanescent operator
does not appear at the one-loop level. 

Combining this result with the previous lattice one we finally 
obtain the relation between the operators ${\cal O}_{PD}^{\msbar}$ 
and ${\cal O}_{PD}^{lat}$:
\begin{eqnarray}
{\cal O}_{PD}^{\msbar}(\mu) & = &\frac{1}{(1-w_0^2)^{3/2} Z_w^{3/2}}
Z_{PD} (\mu a ) {\cal O}_{PD}^{lat} (1/a),
\end{eqnarray}
with
\begin{eqnarray}
Z_{PD} (\mu a) &=& \frac{ (Z_2^{\msbar})^{3/2} V_{PD}^{\msbar}}
{(Z_2)^{3/2} V_{PD}} \nonumber \\
&=& 1 + \frac{g^2}{16\pi^2}\left[
(\delta_{PD} - 3 C_F/2)\log (\mu a)^2 + z_{PD} \right],
 \\
z_{PD} &=& v_{PD}^{\msbar} - v_{PD} + \frac{3}{2} C_F\{\Sigma_1^{\msbar}
-\Sigma_1\}.
\end{eqnarray}
We present numerical values for $z_{PD}$ in 
Table~\ref{tab:total} and
those for the mean-field improved one, $z_{PD}^{MF}$,
in Table~\ref{tab:totalMF}.

\section{Renormalization factor for $B_K$}
\label{sec:bk}

As an application of results in the previous sections, 
we estimate a renormalization factor
for the kaon $B$ parameter $B_K$, defined by
\begin{equation}
B_K =\frac{\langle \overline{K}^0 \vert {\cal O}_+ \vert K^0 \rangle}
{\frac{8}{3} \langle \overline{K}^0 \vert A_4 \vert 0 \rangle
\langle 0 \vert A_4 \vert K^0 \rangle}
\end{equation}
with $q_1=q_3 = s$ and $q_2=q_4= d$ in ${\cal O}_+$.

Denoting the renormalization factor 
between the continuum $B_K$ at scale $\mu$
and the lattice one at scale $1/a$ as $Z_{B_K}( \mu a )$, we obtain
\begin{equation}
Z_{B_K}( \mu a ) =\frac{(1-w_0^2)^{-2} Z_w^{-2} Z_+ (\mu a )}
{(1-w_0)^{-2} Z_w^{-2} Z_A(\mu a)^2}
=\frac{Z_+ (\mu a )}{Z_A(\mu a)^2},
\end{equation}
where
\begin{equation}
Z_+(\mu a) = 1 + \frac{g^2}{16\pi^2}\left[ -4\log (\mu a) + z_+ \right]
\end{equation}
from eq.~(\ref{eq:4fermi}) in this paper, and
\begin{equation}
Z_A(\mu a) = 1 + \frac{g^2 C_F}{16\pi^2} z_A
\end{equation}
from Ref.~\cite{AIKT98}, so that
\begin{equation}
Z_{B_K}(\mu a) = 1 +\frac{g^2}{16\pi^2}\left[ -4\log (\mu a) + z_+ 
-2 C_F z_A \right] .
\end{equation}
Note that $z_A$ in Ref.~\cite{AIKT98} is evaluated in the NDR scheme while
the DRED scheme is used for $z_+$ in this paper.
From the result in Appendix B we have
\begin{equation}
z_A({\rm DRED}) = z_A({\rm NDR}) +1/2, \qquad
z_+({\rm DRED}) = z_+({\rm NDR}) +3 .
\end{equation}

In Ref.\cite{Blum-Soni} $B_K$ has been evaluated at $\beta = 5.85$, 6.0
with $ M= 1.7$ and $\beta = 6.3$ with $M=1.5$, using domain-wall QCD
with the quenched approximation.
Here we explicitly calculate
$Z_{B_K}( \mu a )$ for these parameters. From Table~\ref{tab:total} and 
the previous result\cite{AIKT98}, 
$z_+ = -41.854 (-42.399)$, $z_A = -17.039 (-16.827)$
and $z_+ -2C_F z_A = 3.583 (2.473)$ for $M=1.7 (1.5)$ in the DRED scheme, and
$z_+ = -44.854 (-45.399)$, $z_A = -17.539 (-17.327)$
and $z_+ -2C_F z_A = 1.917 (0.8063)$ for $M=1.7 (1.5)$ in the NDR scheme.
Taking $\mu =1/a$ and
$g^2 = g^2_{\overline{MS}}(1/a)$, estimated by the formula
\begin{equation}
\frac{1}{g^2_{\overline{MS}}}(1/a) = P\frac{\beta}{6} -0.13486
\end{equation}
for the quenched QCD with $P$ being the average value of the plaquette,
we have $Z_{B_K} = $ 1.053 (1.029), 1.049 (1.026) and 1.030 (1.010) 
at $\beta =$ 5.85, 6.0 and 6.3, respectively, in the DRED (NDR) scheme.
Sizes of one-loop corrections for $B_K$
are not so large, $1-5\%$, at these $\beta$ values even without mean-field
improvement,
since the large contribution, which comes from a $(1-w_0) Z_w$ factor,
cancels out in the ratio of ${\cal O}_+$ and $A_4^2$.

If we employ the mean-field improvement by replacing $M\rightarrow
\wt{M}=M+4(u-1)$ with $u = P^{1/4}$, we obtain
$Z_{B_K} = $ 1.018 (0.994), 1.017 (0.994) and 1.009 (0.988) 
at $\beta =$ 5.85, 6.0 and 6.3, respectively, in the DRED (NDR) scheme.
See Appendix A for some remarks.

Note that there is no mean-field improvement factor for $B_K$
in actual simulations since, as mentioned before, it is defined by 
the ratio. Therefore the difference between values of $Z_{B_K}$ 
with and without mean-field improvement comes from
higher order ambiguity in perturbation theory.

Necessary informations for the analysis in this section
are given in Table~\ref{tab:bk}, together with values of $Z_{B_K}$.

\section{Conclusion}
\label{sec:concl}

In this paper we have calculated the one-loop contributions
for the renormalization factors
of the three- and four-quark operators in DWQCD.
We have demonstrated that the three- and four-quark operators 
in DWQCD can be renormalized without any operator mixing between
different chiralities as opposed to the
Wilson case. This desirable property in DWQCD
would practically surpass the cost of the introduction of 
an unphysical fifth dimension.
The numerical values for the finite parts $z_X$ with $X=\pm,1,2,PD$
settle in reasonable magnitude with the mean-field improvement,
while unimproved values are rather large in general.

In this work we do not treat the operators 
which yield the so-called ``penguin'' diagram.
It seems feasible to carry out the calculation of 
their renormalization factors, 
which we leave to future investigation.

\section*{Acknowledgments}

This work is supported in part by the Grants-in-Aid for
Scientific Research from the Ministry of Education, Science and Culture
(Nos. 2373, 2375). 
T. I., Y. K. and Y. T. are supported by Japan Society for Promotion of
Science.

\section*{Appendix A: Mean-field improvement}
The mean-field improvement\cite{MF} in our paper uses
\[
u= 1 - \frac{g^2 C_F}{2}T 
\]
with $T= 0.15493$, which is the value for the link in Feynman gauge.
It may be better to use $u$ from $K_c$ or plaquette in DWQCD. In that case
\[
u = 1 - \frac{g^2 C_F}{2}(T + \delta T),
\]
where $\delta T =0.00793$  for $K_c$, or $\delta T = -0.02993$ for
plaquette. Accordingly we have to modify renormalization factors as follows:
\begin{eqnarray}
z_w^{MF}(T+\delta T) &=& z_w^{MF}(T)+\frac{2w_0}{1-w_0^2}16\pi^2\times
2\delta T, \\
z_2^{MF}(T+\delta T) &=&z_2^{MF}(T)+16\pi^2\times \delta T /2, \\
z_{\Gamma,\rm bilinear}^{MF}(T+\delta T) &=& z_{\Gamma,\rm bilinear}^{MF}(T)
+16\pi^2 \times \delta T/2, \\
z_{\Gamma,\rm 4-quark}^{MF}(T+\delta T) &=& z_{\Gamma,\rm 4-quark}^{MF}(T)
+16\pi^2 \times\delta T/2 \times 2 C_F, \\
z_{\Gamma,\rm 3-quark}^{MF}(T+\delta T) &=& z_{\Gamma,\rm 3-quark}^{MF}(T)
+16\pi^2 \times\delta T/2 \times \frac{3}{2} C_F .
\end{eqnarray}

\section*{Appendix B: Naive Dimensional Regularization(NDR)}
In this appendix we compile the finite part of 
the renormalization constant
in the $\msbar$ subtraction scheme 
with the Naive Dimensional Regularization:
\begin{eqnarray}
\Sigma_1^{\msbar} & =&  1/2, \\
v_+^{\msbar} &=& \delta_+\times\{3/2 - \frac{2N+3}{N-2}\}, \\
v_-^{\msbar} &=& \delta_-\times\{3/2 - \frac{2N-3}{N+2}\}, \\
v_{ij}^{\msbar} & = &
\left(
\begin{array}{cc}
\delta_1 \times\left\{3/2-\displaystyle \frac{2(N^2-5)}{N^2-4}\right\}, & 
\delta_2 \times 3/4 \\
\displaystyle\frac{1}{N}\times 3, & \delta_2\times 1/2
\end{array}
\right), \\
v_{PD}^{\msbar} & = & \delta_{PD}\times 2/3,
\end{eqnarray}
where $v_{ij}$ with $i,j=1,2$ is a matrix, which represents the
mixing of the finite part for ${\cal O}_{1,2}$.
The one-loop vertex corrections for ${\cal O}_{\Gamma}$ 
$(\Gamma=\pm,1,2)$ require to specify their evanescent operators,
which originates from the property that the Fierz transformation
can not be defined in the NDR scheme.
We employ
\begin{eqnarray}
E^{\rm NDR}_{\pm}&=&
\gamma_\rho\gamma_\delta\gamma_\mu(1-\gamma_5)\otimes
\gamma_\mu(1-\gamma_5)\gamma_\delta\gamma_\rho
-{(2-D)^2}\gamma_\mu(1-\gamma_5)\otimes\gamma_\mu(1-\gamma_5),\\
E^{\rm NDR}_{1,2}&=&
\gamma_\rho\gamma_\delta\gamma_\mu(1-\gamma_5)\otimes
\gamma_\mu(1+\gamma_5)\gamma_\delta\gamma_\rho
-{D^2}\gamma_\mu(1-\gamma_5)\otimes\gamma_\mu(1+\gamma_5),
\end{eqnarray}
where $D$ is the reduced space-time dimension.
On the other hand, the evanescent operator does not appear
in the one-loop vertex correction of ${\cal O}_{PD}$.

For later convenience, values of the finite part of quark bilinear operators
are also given here. For NDR scheme
\begin{equation}
z_{V,A}^{\msbar} = 0, \qquad z_{S,P}^{\msbar} = 5/2,
\qquad z_T^{\msbar} = 1/2,
\end{equation}
while for DRED scheme
\begin{equation}
z_{V,A}^{\msbar} = 1/2, \qquad z_{S,P}^{\msbar} = 7/2,
\qquad z_T^{\msbar} = -1/2,
\end{equation}
where the evanescent operators are
\begin{eqnarray}
E^{\rm DRED}_{\gamma_\mu}&=&
{\bar \delta}_{\mu\nu}\gamma_\nu-\frac{D}{4}\gamma_\mu, \\
E^{\rm DRED}_{\gamma_\mu\gamma_5}&=&
{\bar \delta}_{\mu\nu}\gamma_\nu\gamma_5-\frac{D}{4}\gamma_\mu\gamma_5.
\end{eqnarray}

\newcommand{\J}[4]{{\it #1} {\bf #2} (19#3) #4}
\newcommand{\MPL}{Mod.~Phys.~Lett.}
\newcommand{\IJMP}{Int.~J.~Mod.~Phys.}
\newcommand{\NP}{Nucl.~Phys.}
\newcommand{\PL}{Phys.~Lett.}
\newcommand{\PR}{Phys.~Rev.}
\newcommand{\PRL}{Phys.~Rev.~Lett.}
\newcommand{\AP}{Ann.~Phys.}
\newcommand{\CMP}{Commun.~Math.~Phys.}
\newcommand{\PTP}{Prog. Theor. Phys.}
\newcommand{\Suppl}{Prog. Theor. Phys. Suppl.}


\begin{table}
\caption{Color factors for $I_\Gamma^{a,b,c}$ ($\Gamma=\pm,1,2$).}
\label{tab:color}
\begin{center}
\begin{tabular}{llll}
$\Gamma$ & $J_a^{AB}$ & $J_b^{AB}$ & $J_c^{AB}$  \\
\hline
$\pm$ & $T^A T^B \wt{\otimes} 1 \pm T^A \wt{\odot} T^B$
      & $T^A \wt{\otimes} T^B \pm T^A \wt{\odot} T^B$
      & $T^A \wt{\otimes} T^B \pm T^A T^B \wt{\odot} 1$ \\
$1$   & $-N T^A T^B \wt{\otimes} 1 + T^A \wt{\odot} T^B$
      & $-N T^A \wt{\otimes} T^B + T^A \wt{\odot} T^B$
      & $-N T^A \wt{\otimes} T^B + T^A T^B \wt{\odot} 1$ \\
$2$   & $T^A \wt{\odot} T^B$
      & $T^A \wt{\odot} T^B$
      & $T^A T^B \wt{\odot} 1$  \\
\end{tabular}
\end{center}
\end{table}

\begin{table}
\caption{Numerical values for $V_\Gamma$ ($\Gamma=\pm,1,2,PD$) 
as a function of $M$.}
\label{tab:4fermi}
\begin{center}
\begin{tabular}{lllll|l}
$M$ & $V_+$ & $V_- $ & $V_1$ & $V_2$ & $V_{PD}$  \\
\hline
0.05 & 13.9096(8) & 10.847(8) & 13.3992(19) & 8.805(12)& 8.646(5) \\
0.10 & 13.5696 & 11.537 & 13.2309 & 10.182&   8.992 \\
0.15 & 13.2941 & 12.098 & 13.0948 & 11.301&   9.273 \\
0.20 & 13.0548 & 12.587 & 12.9768 & 12.275&   9.518 \\
0.25 & 12.8391 & 13.029 & 12.8708 & 13.155&   9.740 \\
0.30 & 12.6404 & 13.438 & 12.7734 & 13.970&   9.946 \\
0.35 & 12.4542 & 13.822 & 12.6822 & 14.734&  10.139 \\
0.40 & 12.2775 & 14.188 & 12.5960 & 15.462&  10.323 \\
0.45 & 12.1083 & 14.539 & 12.5135 & 16.160&  10.499 \\
0.50 & 11.9449 & 14.880 & 12.4341 & 16.837&  10.671 \\
0.55 & 11.7861 & 15.212 & 12.3571 & 17.496&  10.838 \\
0.60 & 11.6307 & 15.538 & 12.2819 & 18.143&  11.002 \\
0.65 & 11.4779 & 15.859 & 12.2081 & 18.780&  11.164 \\
0.70 & 11.3269 & 16.178 & 12.1354 & 19.412&  11.325 \\
0.75 & 11.1770 & 16.495 & 12.0634 & 20.041&  11.485 \\
0.80 & 11.0275 & 16.813 & 11.9917 & 20.670&  11.645 \\
0.85 & 10.8779 & 17.131 & 11.9201 & 21.300&  11.806 \\
0.90 & 10.7276 & 17.452 & 11.8484 & 21.935&  11.968 \\
0.95 & 10.5760 & 17.777 & 11.7762 & 22.578&  12.133 \\
1.00 & 10.4225 & 18.107 & 11.7033 & 23.230&  12.300 \\
1.05 & 10.2659 & 18.437 & 11.6278 & 23.885&  12.466 \\
1.10 & 10.1076 & 18.790 & 11.5547 & 24.579&  12.646 \\
1.15 & 9.9443 & 19.139 & 11.4768 & 25.269 &  12.822 \\
1.20 & 9.7768 & 19.505 & 11.3981 & 25.990 &  13.007 \\
1.25 & 9.6037 & 19.880 & 11.3165 & 26.732 &  13.198 \\
1.30 & 9.4244 & 20.272 & 11.2323 & 27.504 &  13.396 \\
1.35 & 9.2375 & 20.680 & 11.1446 & 28.309 &  13.603 \\
1.40 & 9.0419 & 21.109 & 11.0530 & 29.153 &  13.820 \\
1.45 & 8.8361 & 21.560 & 10.9567 & 30.042 &  14.049 \\
1.50 & 8.6183 & 22.037 & 10.8547 & 30.983 &  14.291 \\
1.55 & 8.3863 & 22.547 & 10.7464 & 31.987 &  14.550 \\
1.60 & 8.1375 & 23.093 & 10.6301 & 33.063 &  14.827 \\
1.65 & 7.8685 & 23.683 & 10.5043 & 34.227 &  15.127 \\
1.70 & 7.5747 & 24.328 & 10.3668 & 35.496 &  15.454 \\
1.75 & 7.2502 & 25.038 & 10.2149 & 36.897 &  15.814 \\
1.80 & 6.8864 & 25.832 & 10.0440 & 38.463 &  16.216 \\
1.85 & 6.4706 & 26.737 & 9.8482 & 40.247 &   16.675 \\
1.90 & 5.9812 & 27.794 & 9.6168 & 42.337 &   17.210 \\
1.95 & 5.3749 & 29.093 & 9.3281 & 44.907 &   17.867 \\
\end{tabular}
\end{center}
\end{table}

\begin{table}
\caption{Numerical values for $z_\Gamma$ ($\Gamma=\pm,1,2,PD$) 
as a function of $M$.}
\label{tab:total}
\begin{center}
\begin{tabular}{lllll|l}
$M$ &$z_+$ & $z_-$ & $z_1$ & $z_2$ & $z_{PD}$\\
\hline
0.05 & -49.908(10) & -40.845(11) & -48.397(8) & -34.803(15)&-32.144(7)\\
0.10 & -49.332 & -41.300 & -47.994 & -35.945& -32.314  \\
0.15 & -48.847 & -41.651 & -47.647 & -36.853& -32.437  \\
0.20 & -48.416 & -41.949 & -47.338 & -37.637& -32.539  \\
0.25 & -48.025 & -42.215 & -47.057 & -38.341& -32.629  \\
0.30 & -47.663 & -42.461 & -46.796 & -38.993& -32.713  \\
0.35 & -47.325 & -42.693 & -46.553 & -39.605& -32.792  \\
0.40 & -47.007 & -42.918 & -46.326 & -40.191& -32.870  \\
0.45 & -46.706 & -43.137 & -46.111 & -40.758& -32.948  \\
0.50 & -46.419 & -43.354 & -45.908 & -41.311& -33.027  \\
0.55 & -46.145 & -43.571 & -45.716 & -41.855& -33.108  \\
0.60 & -45.883 & -43.790 & -45.534 & -42.395& -33.191  \\
0.65 & -45.631 & -44.012 & -45.361 & -42.933& -33.279  \\
0.70 & -45.388 & -44.239 & -45.196 & -43.473& -33.370  \\
0.75 & -45.153 & -44.472 & -45.040 & -44.017& -33.467  \\
0.80 & -44.927 & -44.712 & -44.891 & -44.569& -33.570  \\
0.85 & -44.708 & -44.961 & -44.750 & -45.130& -33.679  \\
0.90 & -44.496 & -45.220 & -44.617 & -45.704& -33.795  \\
0.95 & -44.290 & -45.491 & -44.490 & -46.292& -33.918  \\
1.00 & -44.091 & -45.776 & -44.372 & -46.899& -34.051  \\
1.05 & -43.898 & -46.070 & -44.260 & -47.517& -34.190  \\
1.10 & -43.710 & -46.392 & -44.157 & -48.181& -34.347  \\
1.15 & -43.528 & -46.723 & -44.061 & -48.853& -34.510  \\
1.20 & -43.352 & -47.079 & -43.973 & -49.565& -34.688  \\
1.25 & -43.180 & -47.457 & -43.893 & -50.308& -34.880  \\
1.30 & -43.014 & -47.862 & -43.822 & -51.094& -35.088  \\
1.35 & -42.853 & -48.296 & -43.760 & -51.924& -35.315  \\
1.40 & -42.697 & -48.763 & -43.708 & -52.808& -35.561  \\
1.45 & -42.545 & -49.269 & -43.666 & -53.751& -35.831  \\
1.50 & -42.399 & -49.818 & -43.635 & -54.763& -36.127  \\
1.55 & -42.257 & -50.417 & -43.617 & -55.857& -36.453  \\
1.60 & -42.119 & -51.074 & -43.611 & -57.044& -36.813  \\
1.65 & -41.985 & -51.800 & -43.621 & -58.343& -37.214  \\
1.70 & -41.854 & -52.607 & -43.646 & -59.776& -37.663  \\
1.75 & -41.725 & -53.513 & -43.690 & -61.372& -38.170  \\
1.80 & -41.595 & -54.541 & -43.753 & -63.172& -38.748  \\
1.85 & -41.460 & -55.727 & -43.838 & -65.237& -39.417  \\
1.90 & -41.311 & -57.124 & -43.946 & -67.666& -40.207  \\
1.95 & -41.121 & -58.840 & -44.074 & -70.652& -41.177  \\
\end{tabular}
\end{center}
\end{table}

\begin{table}
\caption{Numerical value for $z_\Gamma^{MF}$ ($\Gamma=\pm,1,2,PD$)  
as a function of $M$.}
\label{tab:totalMF}
\begin{center}
\begin{tabular}{lllll|l}
$M$ &$z_+^{MF}$ & $z_-^{MF}$ & $z_1^{MF}$ & $z_2^{MF}$& $z_{PD}^{MF}$ \\
\hline
0.05 & -17.287(10) & -8.224(11) & -15.776(8) & -2.182(15)& -7.679(7) \\
0.10 & -16.712 & -8.679 & -15.373 & -3.324&  -7.848 \\
0.15 & -16.226 & -9.030 & -15.027 & -4.232&  -7.972 \\
0.20 & -15.796 & -9.328 & -14.718 & -5.016&  -8.074 \\
0.25 & -15.404 & -9.594 & -14.436 & -5.720&  -8.164 \\
0.30 & -15.042 & -9.840 & -14.175 & -6.372&  -8.247 \\
0.35 & -14.705 & -10.073 & -13.933 & -6.98&  -8.326  \\
0.40 & -14.386 & -10.297 & -13.705 & -7.57&  -8.404  \\
0.45 & -14.085 & -10.516 & -13.490 & -8.13&  -8.482  \\
0.50 & -13.799 & -10.734 & -13.288 & -8.69&  -8.561  \\
0.55 & -13.525 & -10.951 & -13.096 & -9.23&  -8.642  \\
0.60 & -13.262 & -11.169 & -12.913 & -9.77&  -8.726  \\
0.65 & -13.010 & -11.391 & -12.740 & -10.3&  -8.813  \\
0.70 & -12.767 & -11.618 & -12.575 & -10.8&  -8.905  \\
0.75 & -12.532 & -11.851 & -12.419 & -11.3&  -9.002  \\
0.80 & -12.306 & -12.091 & -12.270 & -11.9&  -9.104  \\
0.85 & -12.087 & -12.340 & -12.129 & -12.5&  -9.213  \\
0.90 & -11.875 & -12.600 & -11.996 & -13.0&  -9.329  \\
0.95 & -11.670 & -12.871 & -11.870 & -13.6&  -9.453  \\
1.00 & -11.470 & -13.155 & -11.751 & -14.2&  -9.585  \\
1.05 & -11.278 & -13.449 & -11.639 & -14.8&  -9.725  \\
1.10 & -11.089 & -13.772 & -11.536 & -15.5&  -9.882  \\
1.15 & -10.908 & -14.103 & -11.440 & -16.2& -10.044  \\
1.20 & -10.731 & -14.459 & -11.352 & -16.9& -10.223  \\
1.25 & -10.559 & -14.836 & -11.272 & -17.6& -10.414  \\
1.30 & -10.393 & -15.241 & -11.201 & -18.4& -10.623  \\
1.35 & -10.232 & -15.675 & -11.139 & -19.3& -10.849  \\
1.40 & -10.076 & -16.143 & -11.087 & -20.1& -11.096  \\
1.45 & -9.925 & -16.648 & -11.045 & -21.13& -11.365  \\
1.50 & -9.778 & -17.197 & -11.014 & -22.14& -11.661  \\
1.55 & -9.636 & -17.796 & -10.996 & -23.23& -11.987  \\
1.60 & -9.498 & -18.453 & -10.991 & -24.42& -12.347  \\
1.65 & -9.364 & -19.179 & -11.000 & -25.72& -12.749  \\
1.70 & -9.233 & -19.986 & -11.025 & -27.15& -13.198  \\
1.75 & -9.104 & -20.892 & -11.069 & -28.75& -13.705  \\
1.80 & -8.975 & -21.920 & -11.132 & -30.55& -14.283  \\
1.85 & -8.840 & -23.106 & -11.217 & -32.61& -14.952  \\
1.90 & -8.690 & -24.503 & -11.326 & -35.04& -15.742  \\
1.95 & -8.500 & -26.219 & -11.453 & -38.03& -16.711  \\
\end{tabular}
\end{center}
\end{table}

\begin{table}
\caption{Renormalization factor for $B_K$($1/a$) at some parameters.}
\label{tab:bk}
\begin{center}
\begin{tabular}{l|ll|ll|ll}
$\beta$ &\multicolumn{2}{c|}{5.85} & \multicolumn{2}{c|}{6.0} & 
\multicolumn{2}{c}{6.3} \\
$M$     & \multicolumn{2}{c|}{1.70} &\multicolumn{2}{c|}{1.70} &
\multicolumn{2}{c}{1.50} \\
\hline
$P$ & \multicolumn{2}{c|}{0.57506} & \multicolumn{2}{c|}{0.59374} & 
\multicolumn{2}{c}{0.62246} \\
$g^2_{\overline{MS}}(1/a)$ & \multicolumn{2}{c|}{2.3484} & 
\multicolumn{2}{c|}{2.1792} & \multicolumn{2}{c}{1.9278} \\
$u$ & \multicolumn{2}{c|}{0.87082} & \multicolumn{2}{c|}{0.87781} & 
\multicolumn{2}{c}{0.88823} \\
$\wt{M}$ & \multicolumn{2}{c|}{1.20} & \multicolumn{2}{c|}{1.20} & 
\multicolumn{2}{c}{1.05} \\
\hline
 & DRED & NDR & DRED & NDR & DRED & NDR \\
\hline
$z_+$ & -41.854 & -44.854 & -41.854 & -44.854 & -42.399 & -45.399 \\
$z_A$ & -17.039 & -17.539 & -17.039 & -17.539 & -16.827 & -17.327 \\
$z_+ - 2C_F z_A$ & 3.583 & 1.917 & 3.583 & 1.917 & 2.473 & 0.806 \\
$Z_{B_K}(\mu a=1)$ & 1.053 & 1.029 & 1.049 & 1.026 & 1.030 & 1.010 \\
\hline
$z_+^{MF}$ & -17.033 & -20.033 & -17.033 & -20.033 & -17.580 & -20.580 \\
$z_A^{MF}$ & -6.853 & -7.353 & -6.853 & -7.353 & -6.864 & -7.364 \\
$z_+^{MF} - 2C_F z_A^{MF}$ & 1.242& -0.425 & 1.242 & -0.425 & 0.724 & -0.943 \\
$Z_{B_K}^{MF}(\mu a=1)$ & 1.018 & 0.994 & 1.017 & 0.994 & 1.009 & 0.988 \\
\end{tabular}
\end{center}
\end{table}
\begin{figure}
\begin{center}
\begin{picture}(420,450)(0,20)
\Text(60,60)[l]{${\rm a}^\prime$}
\Vertex(70,166){4.5}
\Vertex(70,154){4.5}
\ArrowLine(68,166)(8,226)
\ArrowLine(132,226)(72,166)
\ArrowLine(68,154)(8,94)
\ArrowLine(132,94)(72,154)
\Gluon(20,105)(120,105){5}{8}
%
\Text(60,260)[l]{a}
\Text(0,440)[l]{$\alpha,i$}
\Text(115,440)[l]{$\beta,j$}
\Text(0,285)[l]{$\gamma,k$}
\Text(115,285)[l]{$\delta,l$}
\Vertex(70,366){4.5}
\Vertex(70,354){4.5}
\ArrowLine(68,366)(8,426)
\ArrowLine(132,426)(72,366)
\ArrowLine(68,354)(8,294)
\ArrowLine(132,294)(72,354)
\Gluon(20,415)(120,415){-5}{8}
%
\Text(200,60)[l]{${\rm b}^\prime$}
\Vertex(210,166){4.5}
\Vertex(210,154){4.5}
\ArrowLine(208,166)(148,226)
\ArrowLine(272,226)(212,166)
\ArrowLine(208,154)(148,94)
\ArrowLine(272,94)(212,154)
\Gluon(260,213)(260,107){-5}{8}
%
\Text(200,260)[l]{b}
\Vertex(210,366){4.5}
\Vertex(210,354){4.5}
\ArrowLine(208,366)(148,426)
\ArrowLine(272,426)(212,366)
\ArrowLine(208,354)(148,294)
\ArrowLine(272,294)(212,354)
\Gluon(160,413)(160,307){5}{8}
%
\Text(340,60)[l]{${\rm c}^\prime$}
\Vertex(350,166){4.5}
\Vertex(350,154){4.5}
\ArrowLine(348,166)(288,226)
\ArrowLine(412,226)(352,166)
\ArrowLine(348,154)(288,94)
\ArrowLine(412,94)(352,154)
\GlueArc(350,160)(30,51,230){-5}{7}
%
\Text(340,260)[l]{c}
\Vertex(350,366){4.5}
\Vertex(350,354){4.5}
\ArrowLine(348,366)(288,426)
\ArrowLine(412,426)(352,366)
\ArrowLine(348,354)(288,294)
\ArrowLine(412,294)(352,354)
\GlueArc(350,360)(30,-50,130){-5}{7}
\end{picture}
\end{center}
\caption{One-loop vertex corrections for the four-quark operator.
$\alpha,\beta,\gamma,\delta$ and $i,j,k,l$ label Dirac and color
indices respectively.}
\label{fig:vc_4}
\end{figure}
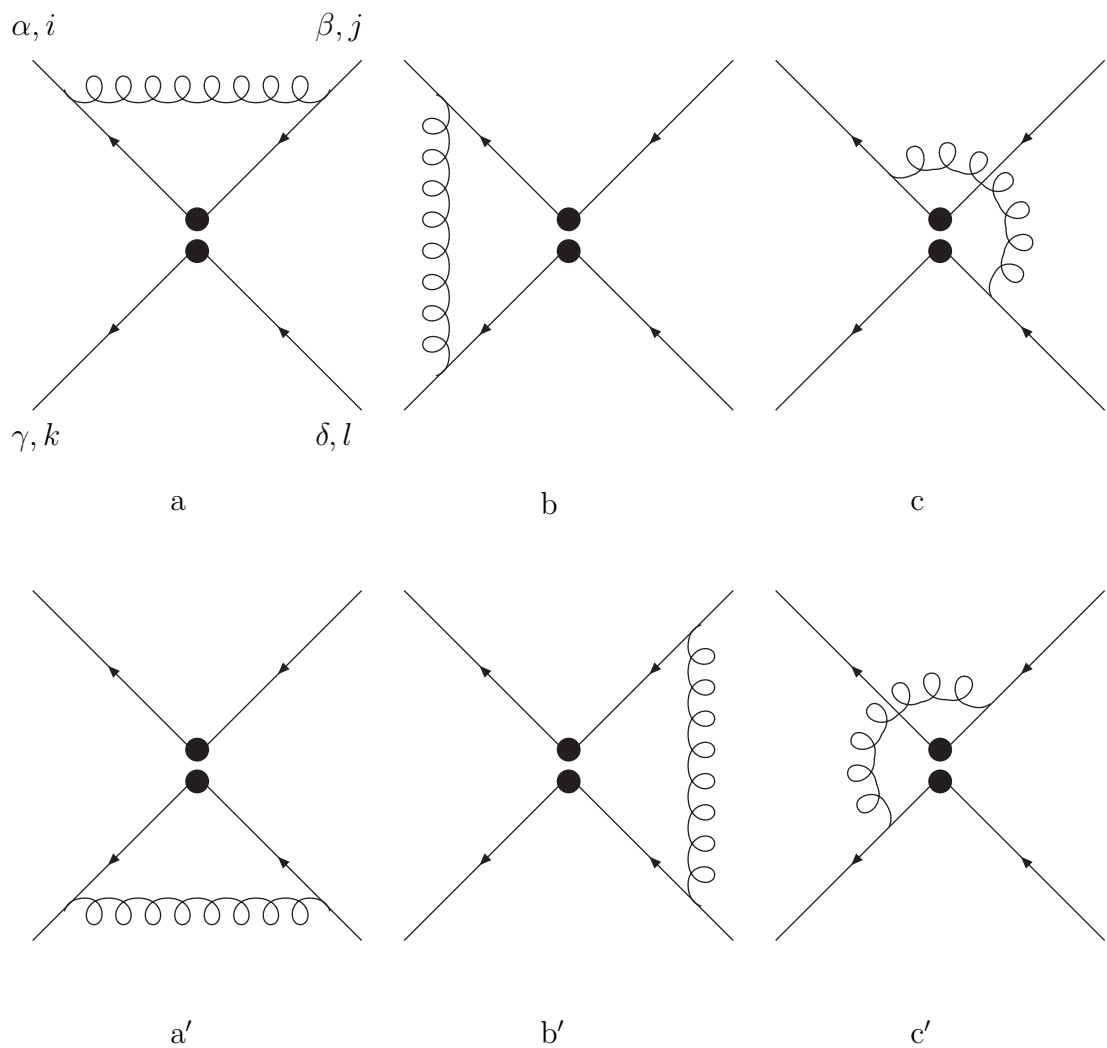

\begin{figure}
\begin{center}
\begin{picture}(420,250)(0,20)
\Text(60,60)[l]{a}
\Text(0,240)[l]{$\alpha,i$}
\Text(115,240)[l]{$\beta,j$}
\Text(115,85)[l]{$\gamma,k$}
\Vertex(70,166){4.5}
\Vertex(70,154){4.5}
\ArrowLine(68,166)(8,226)
\ArrowLine(132,226)(72,166)
\ArrowLine(132,94)(72,154)
\Gluon(20,215)(120,215){-5}{8}
%
\Text(200,60)[l]{b}
\Vertex(210,166){4.5}
\Vertex(210,154){4.5}
\ArrowLine(208,166)(148,226)
\ArrowLine(272,226)(212,166)
\ArrowLine(272,94)(212,154)
\Gluon(260,213)(260,107){-5}{8}
%
\Text(340,60)[l]{c}
\Vertex(350,166){4.5}
\Vertex(350,154){4.5}
\ArrowLine(348,166)(288,226)
\ArrowLine(412,226)(352,166)
\ArrowLine(412,94)(352,154)
\GlueArc(350,160)(30,-50,130){-5}{7}
\end{picture}
\end{center}
\caption{One-loop vertex corrections for the three-quark operator.
$\alpha,\beta,\gamma$ and $i,j,k$ label Dirac and color
indices respectively.}
\label{fig:vc_3}
\end{figure}
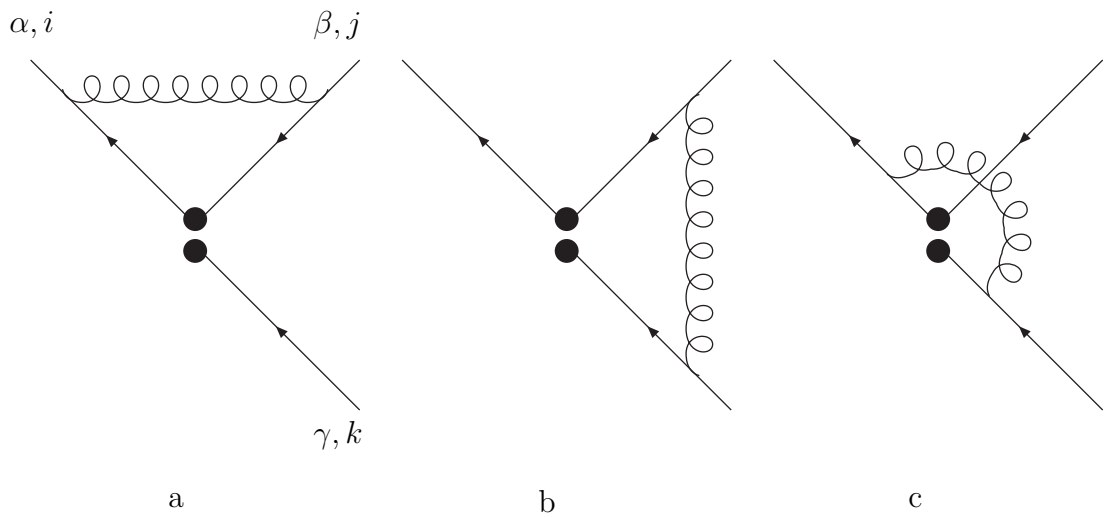

\end{document}